\documentclass[aps,pre,twocolumn,groupedaddress]{revtex4-2}

\usepackage{amsmath,amssymb,bm,graphicx}
\usepackage{hyperref}
\usepackage{enumitem}

\DeclareMathOperator{\Var}{Var}

\begin{document}

\title{A generative model for bipartite gene-sharing networks}

\author{Jaime Iranzo}
\affiliation{Centro de Astrobiolog\'{\i}a (CSIC-INTA), 28864 Madrid, Spain}
\affiliation{Grupo Interdisciplinar de Sistemas Complejos (GISC), 28911 Legan\'es, Spain}
\affiliation{Instituto de Biocomputaci\'on y F\'{\i}sica de Sistemas Complejos (BIFI), Universidad de Zaragoza, 50018 Zaragoza, Spain}

\author{Pedro J\'odar}
\affiliation{Universidad Carlos III de Madrid, Departamento de Matem\'aticas, 28911 Legan\'es, Spain}
\affiliation{Grupo Interdisciplinar de Sistemas Complejos (GISC), 28911 Legan\'es, Spain}

\author{Eugene V. Koonin}
\affiliation{Computational Biology Branch, Division of Intramural Research, National Library of Medicine, National Institutes of Health, Bethesda, MD 20894, USA}

\author{Susanna Manrubia}
\affiliation{Department of Biodiversity and Evolutionary Biology, Museo Nacional de Ciencias Naturales (CSIC), 28006 Madrid, Spain}
\affiliation{Grupo Interdisciplinar de Sistemas Complejos (GISC), 28911 Legan\'es, Spain}

\author{Jos\'e A. Cuesta}
\email{cuesta@math.uc3m.es}
\affiliation{Universidad Carlos III de Madrid, Departamento de Matem\'aticas, 28911 Legan\'es, Spain}
\affiliation{Grupo Interdisciplinar de Sistemas Complejos (GISC), 28911 Legan\'es, Spain}
\affiliation{Instituto de Biocomputaci\'on y F\'{\i}sica de Sistemas Complejos (BIFI), Universidad de Zaragoza, 50018 Zaragoza, Spain}

\date{\today}

\begin{abstract}
Gene-sharing networks provide a powerful framework to study the evolution of viruses and mobile genetic elements. These bipartite networks, which link genes to the genomes that contain them, exhibit characteristic degree distributions: a scale-free distribution for genes and an exponential-like decay for genomes. Here, we propose a mechanistic model that explains these patterns through fundamental evolutionary processes including horizontal gene transfer, capture of new genes, emergence of new genomes, and gene loss. Using a mean-field approximation, we derive analytical expressions for the asymptotic gene and genome degree distributions, recapitulating a power-law distribution for genes and an exponential distribution for genomes. Numerical simulations validate these predictions and yield parameter values that closely fit empirical data from dsDNA viruses, RNA viruses, and prokaryotic pangenomes. This simple model with only two parameters provides a generative framework for bipartite gene-sharing networks, offering qualitative and quantitative insights into the main evolutionary forces driving genome plasticity. Setting the gene loss rate to zero, the gene and genome degree distributions of the model closely fit the empirically observed distributions. Thus, evolution of viruses appears to be dominated by gene gain, in agreement with the results of independent reconstructions of viral evolution.
\end{abstract}

\maketitle

\section{Introduction}

Evolution of the genomes of prokaryotes and viruses is a highly dynamic process that is shaped by extensive gene gain, primarily, via horizontal gene transfer (HGT), and gene loss \cite{doolittle:1999, kunin:2005, koonin:2008, puigbo:2014, booth:2016}. %PMID: 10381871 PMID: 10381871 (repetido) PMID: 15965028 PMID: 18948295 PMID: 25141959 PMID: 27482743
Gene-sharing networks have proven a useful framework to study these evolutionary processes. Unlike traditional phylogenetic approaches, gene-sharing networks do not rely on high-quality alignments of universal marker genes and therefore are particularly useful for investigating deep evolutionary connections among viruses and mobile genetic elements (MGE) \cite{iranzo:2016a,iranzo:2016b,wolf:2018}. %[PMID: 27486193;  27681128; 30482837]
For many practical applications, gene-sharing networks take the form of bipartite graphs \cite{pavlopoulos:2018}, that is, graphs with two classes of nodes, one representing families of orthologous genes and the other representing genomes \cite{iranzo:2017a}. %[PMID: 28451057]
In bipartite gene-sharing networks, links connect each genome with all the gene families present in that genome, and consequently, each gene family is connected with all the genomes in which it is represented. Compared to other representations of cross-genome similarity, bipartite networks preserve all available information about gene sharing patterns. As a result, they can be used to infer gene exchange across genomes, identify conserved subgenomic elements, such as defense systems or transposons, and classify genomes based on their gene content \cite{corel:2018}. %[PMID: 29346651]

Bipartite gene-sharing networks in DNA and RNA viruses are characterized by two major structural properties: modularity and hierarchy \cite{iranzo:2016a}. %[PMID: 27486193]
The modules of the network correspond to groups of viruses that share significantly more genes among themselves than they share with viruses outside the given module. From an evolutionary perspective, shared genes within a module most likely result from common ancestry, although some homologs might have been acquired by different viruses independently. The hierarchical organization of the network likely reflects a tree-like trend in virus diversification, which is detectable despite pervasive gene exchange. In support of this interpretation, the modules of the gene-sharing network agree well with the taxonomy of viruses that was constructed, primarily, on the basis of combined phylogenies of viral hallmark genes, with the lowest level of modules roughly corresponding to viral genera \cite{lima-mendez:2017,wolf:2018}. %[PMID: 18234706; 30482837]
This correspondence between the structure of gene-sharing networks and viral taxonomy has fueled the development of network-based tools for unsupervised virus classification, which are especially crucial and promising to deal with huge numbers of viral genomes assembled from metagenomic data \cite{jang:2019}. %[PMID: 31061483]
Besides these applications, bipartite gene-sharing networks provide an opportunity to investigate the forces and trends underlying the evolution of viral genomes. Network architecture and the dynamical processes that shape it are deeply interconnected \cite{newman:2003}. In the case of bipartite gene-sharing networks, statistical properties of the network, such as the node degree distributions  of genomes and gene families, contain information about the relative rates of gene gain and loss, and the rates of emergence and extinction of viral species as well as higher taxa mapping to the network modules.

The idea of examining gene frequencies to discriminate between different models of genome evolution dates back, at least, to the seminal works on transposon dynamics conducted in the 1980's and early 1990's (e.g. \cite{charlesworth:1983,langley:1983,moody:1988,basten:1991}). %DOI:10.1017/S0016672300021455; PMID: 17246142; 2842425; 1658178]
Since then, the systematic study of gene family sizes within and across prokaryotic and eukaryotic genomes has shed light on fundamental aspects of gene and genome evolution. In particular, by fitting mathematical models of gene diversification to empirical distributions of gene family sizes, Huynen and Van Nimwegen \cite{huynen:1998}, and subsequently, in a refined analysis, Karev et al. \cite{karev:2002,karev:2004} %[PMID: 12379152; 15357876]
provided evidence of a balance between domain innovation, duplication, and loss rates, together with self-accelerated gene diversification. Regarding genome size and composition, generalized birth-and-death models have been combined with empirical data to demonstrate that genome shrinking is a non-adaptive trend in most prokaryotic lineages and the long-term maintenance of slightly beneficial genes requires a minimum rate of horizontal gene transfer \cite{sela:2016,iranzo:2016c,iranzo:2017b}. %[PMID: 27702904; 27503291; 28652353]

Empirical gene-sharing networks display a number of appealing regularities suggestive of dominating, underlying generative processes. Both for RNA and DNA viruses, the gene-degree distribution (that is, the number of genomes that contain a given gene family) is compatible with a power-law distribution, whereas the genome-degree distribution (the number of gene families represented in a given genome) is a fast-decaying, exponential-like function \cite{iranzo:2016a}. In all cases analyzed, gene-sharing networks display a modular structure that recapitulates viral taxa \cite{iranzo:2016a,iranzo:2016b,wolf:2018,jang:2019}. 

Although power-law distributions for gene and domain family sizes have been widely reported and analyzed \cite{huynen:1998,koonin:2002,reed:2004,hugues:2008}, %[PMID: 12432406]
models of gene family expansion able to generate such power laws do not generally consider the concomitant dynamics of genome size. The sizes of gene families and genomes are intimately related, given that genomes are the only environment in which individual genes can thrive. Genomes better represent organisms than any particular gene family and are subject to different selective pressures that, in turn, should affect the evolutionary dynamics of genes and gene families. Popular network models that generate power-law degree distributions, such as the Barabási-Albert preferential attachment model \cite{barabasi:1999, newman:2010}, assume an infinite external pool of isolated elements that are sequentially linked to preexisting elements of the same kind (monopartite networks). Generative models for bipartite networks have not been explored in detail and, to the best of our knowledge, currently, there is no generative model applicable to gene-sharing networks. 

Here we jointly model the evolution of genome and gene family sizes. Our goal is to develop generative models for bipartite networks with properties compatible with those observed in the virus gene-sharing networks. By studying genome size and gene spread under the same unified framework, we conclude that, at odds with genomes of cellular life forms, the structure of viral gene-sharing networks implies that gene gain dominates over gene loss, a conclusion that is compatible with empirical evolutionary reconstructions for several major groups of viruses. We extend the model to prokaryotic pangenomes and show that accessory gene-sharing networks are also subject to an expansion bias. Finally, we demonstrate that genome expansion is not an artifact of the bipartite generative network model.

\section{Mechanistic model}

The mechanistic model we devised includes fundamental processes that are well documented in viral evolution:

\paragraph*{Horizontal gene transfer (HGT)} Genes in the network can be transferred from genome to genome. The higher the abundance of a gene among genomes, the greater the probability that such a gene undergoes HGT, so it seems plausible to assume the likelihood of such an event to be proportional to the abundance of the gene.

\paragraph*{Functional innovation (FI)} New genes, originated in an external and---to all purposes---unlimited pool, can be incorporated into a genome via HGT.

\paragraph*{Organismal innovation (OI)} With a certain probability, a gene undergoing HGT founds a new genome, rather than joining an existing one. Although the birth of a new genome from a single gene may appear an oversimplification, it becomes reasonable under the assumption that the rates of gain, loss, and exchange of core genes are negligible and, therefore, these genes do not fall within the scope of the model. Thus, every genome includes a set of essential genes for which evolution is not explicitly modeled (for example, the conserved morphogenetic toolbox in DNA viruses or the RNA-dependent RNA polymerase in RNA viruses). Accordingly, the emergence of a new genome can be perceived as capture of a new gene on top of that immutable core.

\paragraph*{Gene loss (GL)} Genomes lose genes at a constant rate. Gene loss can result from the accumulation of point mutations and small indels, leading to pseudogenization and, eventually, genome erosion, and also from recombination events leading to large deletions. For the purpose of this model, we combined both mechanisms under a single overall gene loss rate. %This event is very rare though, because it implies the disappearance of one gene from a whole viral species.

\begin{figure*}[tbp!]
\centering
\includegraphics[width=0.9\textwidth]{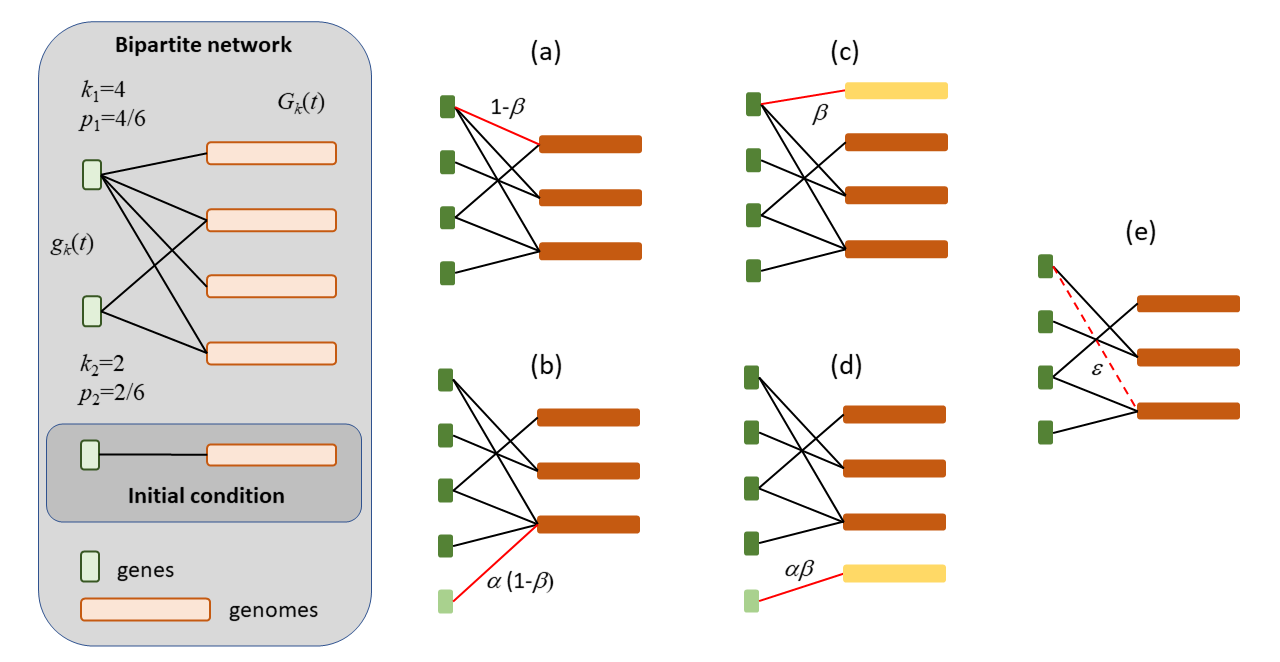}
\caption[]{Sketch of the bipartite network and main quantities and mechanisms included in the gene-sharing model. Left: The degree $k_i$ of a gene is the number of genomes that contain gene $i$ (examples are shown for two genes in the network), while the degree of a genome (its size) corresponds to its number of genes. The probability that a gene or a genome has degree $k$ at time $t$ is given by distributions $g_k(t)$ and $G_k(t)$, respectively. The initial condition of the process is the simplest bipartite network, consisting of one gene, one genome, and a single link between them. Panels (a)-(e) illustrate the possible events occurring at each time step, which mimic the fundamental processes shaping viral genome evolution. (a) pure HGT with gene and genome conservation; (b) addition of a new gene from an external pool to the existing ensemble of genomes; (c) organismal innovation through a new combination of pre-existing genes; (d)  organismal innovation via incorporation of novel genes; and (e) gene loss, which  occurs independently of genome age or size and may occasionally reduce the total number of genes or genomes.}
\label{fig:ModelScheme}
\end{figure*}

\vspace*{3mm}\noindent
These processes can be translated into a set of dynamical rules that generate a bipartite network (Figure~\ref{fig:ModelScheme}). The network emerges through the following dynamical processes:
\begin{enumerate}[label=(\arabic*)]
\item At a certain rate---which we set to 1 to fix the time scale---a random gene in the network undergoes HGT. Among all of them, gene $i$ is selected with a probability $p_i$ proportional to its degree $k_i$, namely, $p_i=k_i/\sum_j k_j$ (Fig.~\ref{fig:ModelScheme}a, c). 

\item With a probability $\alpha$, a new gene from the external pool is added to the system (Fig.~\ref{fig:ModelScheme}b, d). 

\item Regardless of whether the gene is old (process 1) or new (process 2), with probability $1-\beta$ it gets linked to a randomly picked genome (Fig.~\ref{fig:ModelScheme}a, b) or founds a new genome with probability $\beta$; by model construction, in the latter case the new genome will contain this gene only (Fig.~\ref{fig:ModelScheme}c, d). In the former case, it may happen that the selected gene is already present in the selected genome (they are already linked); then, this step is skipped.
%The event depicted in Fig. \ref{fig:ModelScheme}a would be ignored if the link with the chosen genome already existed.

\item With probability $\epsilon$, one link of the network chosen at random is removed, Fig.~\ref{fig:ModelScheme}e. This represents a genome losing a gene. If, as a result, the gene and/or genome involved lose all their links, they are removed from the network.
\end{enumerate}

\section{Mean-field approximation}

Although the model above is easy to implement and simulate computationally, a comprehensive study of the parameters involved is costly. Therefore, before attempting systematic, numerical explorations of the parameter space, we sought to gain insight into the gene-sharing network evolution through a formal, mean-field approximation. This analysis sheds light on the general functional forms of the gene and genome degree distributions, and provides for deriving algebraic relationships between model parameters and average quantities measured in the various data sets that will be subsequently analyzed. To further simplify the analytical treatment, we initially set $\epsilon=0$ and subsequently explore numerically the effect of a positive deletion rate. The consistency of this assumption will be assessed later, and an upper bound for $\epsilon$ will be derived from numerical simulations.

\subsection{Network growth}

The number of nodes in the bipartite network increases at a constant rate: $\alpha$ for genes, $\beta(1+\alpha)$ for genomes. Thus, on average, the number of genes and the number of genomes at time step $t$, are given by $N_g(t)=N_g(0)+\alpha t$, and $N_G(t)=N_G(0)+\beta(1+\alpha)t$, respectively. We will henceforth assume $N_g(0)=N_G(0)=1$ for simplicity.

At every time step new links can be created through the processes described in  the previous section. The addition of a new gene adds a link at a rate $\alpha$, regardless of whether that gene joins a pre-existing genome (Fig.~\ref{fig:ModelScheme}b) or a new genome (Fig.~\ref{fig:ModelScheme}d); a pre-existing gene links to a new genome at a rate $\beta$ (Fig.~\ref{fig:ModelScheme}c). In contrast, linking existing genes to existing genomes (Fig.~\ref{fig:ModelScheme}a) does not add new links at a constant rate because the outcome of such events depends on whether the gene is already present in the chosen genome. For the time being, let us call $P_f(t)$ the probability that the link is formed, so that the rate of creation of new links through this process is $(1-\beta) P_f(t)$. Taking the sum of all possibilities, the mean number of links $L(t)$ at time step $t$ fulfills 
\begin{equation}
    \frac{dL}{dt}=\alpha+\beta+(1-\beta)P_f(t),
    \label{eq:linkgrowth}
\end{equation}
with $L(0)=1$.
Assuming $P_f(t)\to P_{\infty}$ as $t\to\infty$, the long-term solution to this equation behaves as
\begin{equation}
    L(t)\sim\big[\alpha+\beta+(1-\beta)P_{\infty}\big]t.
\label{eq:linklinear}
\end{equation}
The precise value of $P_{\infty}$ will be explicitly obtained later.

\subsection{Degree distribution of genes}

Let us call $g_k(t)$ the probability that a gene is contained in $k$ genomes at time $t$ (that is, the probability that a gene node has degree $k$ in the bipartite network). To derive the master equation for this function, we need to estimate the probability that a gene of degree $k$ forms a new link in the next time step. For this to happen, the gene has to be chosen, which occurs with a probability proportional to $kg_k(t)$, i.e., the number of links belonging to genes of degree $k$, normalized over $L(t)=\sum_kkg_k(t)$. Then, with probability $\beta$, this gene creates a new genome, and with probability $1-\beta$ it links to an extant genome, provided the gene is not already contained in it. Because the gene is already present in $k$ genomes out of $N_G(t)$, the mean-field estimate for the probability of the latter event is $1-k/N_G(t)$. Finally, new genes with one link emerge at a rate $\alpha$. The terms corresponding to all these processes are included in the master equation ($k\ge 1$)
\begin{equation}
    \dot g_k=\left(\mathbb{E}^{-1}-1\right)\frac{k}{L(t)}
    \left[1-(1-\beta)\frac{k}{N_G(t)}\right]g_k
    +\alpha\delta_{k,1},
    \label{eq:master}
\end{equation}
using the \emph{shift} operator $\mathbb{E}$, defined as $\mathbb{E}^sf(k)\equiv f(k+s)$ for every integer $s$; $\delta_{k,j}=1$ if $k=j$ and $0$ otherwise. The presence of the operator $\mathbb{E}^{-1}-1$ means that every time a gene of degree $k-1$ creates a new link, $g_{k-1}(t)$ decreases by one and $g_k(t)$ increases by one.

From this equation we can obtain (see `Materials and methods') the mean-field estimate of the probability of link formation (that is, the probability that a gene is not present in the genome that has been selected for the gene to link to)
\begin{equation}
    P_f(t)=1-\frac{1}{L(t)N_G(t)}\sum_{k\ge 1}k^2g_k(t).
    \label{eq:Pf}
\end{equation}

%\subsubsection{Long-term limit of the gene degree distribution}

For sufficiently small deletion rate $\epsilon$,  the number of genes, genomes and links in the network grows indefinitely, so the system does not reach a stationary state. However, it can be expected that $g_k(t)\sim N_g(t)p_k$, as $t\to\infty$, with $p_k$ independent of $t$. In this limit, \eqref{eq:Pf} yields $P_{\infty}=1$ implying that, after a transient, the event represented in Fig.~\ref{fig:ModelScheme}a is always successful. On the other hand, in this same limit the master \eqref{eq:master} leads to (see `Materials and Methods')
\begin{equation}
\begin{split}
    p_k &=(1+\alpha)\Gamma(2+\alpha)\frac{(k-1)!}{\Gamma(k+2+\alpha)} \\
    &\sim(1+\alpha)\Gamma(2+\alpha)k^{-2-\alpha} \quad (k\to\infty).
\end{split}\label{eq:pkfinal}
\end{equation}
The gene degree distribution decays as a scale-free, power law function with exponent $2+\alpha$. This observation allows us to obtain $\alpha$ from empirical data.

The long-term limit of the gene degree distribution corresponds to Yule's \cite{yule:1925} (or Yule-Simon's \cite{simon:1955}) distribution, \eqref{eq:pkfinal}. Yule originally derived it in a continuous-time birth process to explain the distribution of the number of species per genus. In our framework, the number of links grows as $dL/dt=1+\alpha$, so Yule's model applies directly in the long-time limit. As in Yule's original formulation, transient effects depend on the initial condition; such transients have been analyzed in related birth-death models for family-size distributions 
\cite{manrubia:2002}. The same distribution is also recovered for specific parameter choices in the Barabási-Albert model of preferential attachment \cite{barabasi:1999}. 

Yule's distribution, \eqref{eq:pkfinal}, yields an asymptotic average degree 
\begin{equation*}
\langle k \rangle_g = \frac{1+\alpha}{\alpha}
\end{equation*}
for any $\alpha>0$. The variance of the distribution, however, is finite only for $\alpha>1$. Accordingly, for $0<\alpha<1$, large fluctuations in the average value of the gene degree for any finite system should be expected. Note that a finite average degree is consistent with $P_{\infty}=1$, given that the number of genes and genomes grows linearly with time. Also, the fact that the ratio  

\begin{equation*}
\frac{N_g(t)}{N_G(t)} \to \frac{\alpha}{\beta (1+\alpha)} 
\end{equation*}
becomes asymptotically independent of time indirectly supports our proposition regarding a stationary functional shape for the degree distributions (and, by extension, for any other structural property of the bipartite network). 

\subsection{Degree distribution of genomes}

The master equation for the degree distribution of genomes is easier to derive. If $G_k(t)$ denotes the number of genomes that have degree $k$ at time $t$, the probability that one such genome is chosen for HGT is simply $G_k(t)/N_G(t)$. As the rate at which new links are formed is $\alpha+\beta+(1-\beta)P_f(t)$, and new genomes are founded at rate $\beta (1+\alpha)$, the master equation turns out to be ($k\ge 1$)
\begin{equation*}
    \dot G_k=\frac{\alpha+\beta+(1-\beta)P_f(t)}{N_G(t)}\left(\mathbb{E}^{-1}-1\right)G_k+\beta (1+\alpha)\delta_{k,1},
    %\label{eq:mastergenomes}
\end{equation*}
with the assumption $G_0(t)\equiv 0$.

As $P_f(t)\to 1$ when $t\to\infty$, assuming (as in the case of genes), that in this limit $G_k(t)\sim N_G(t)q_k$, with $q_k$ independent of time, the master equation becomes a simple iteration,
%\begin{align*}
%    &\beta(1+\alpha)q_1=\beta (1+\alpha) -(1+\alpha)q_1, \\
%    &\beta(1+\alpha)q_k=(1+\alpha)(q_{k-1}-q_k),
%\end{align*}
%or better
%\begin{equation*}
%    q_1=\frac{\beta}{1+\beta}, \qquad q_k=\frac{1}{1+\beta} \, q_{k-1},
%\end{equation*}
whose solution is just
\begin{equation}
    q_k=\beta\left(\frac{1}{1+\beta}\right)^k.
    \label{eq:qkfinal}
\end{equation}
The value of $\beta$ follows from fitting the exponential distribution to empirical data. In the mean-field approximation, the asymptotic degree distribution of genomes is a decaying exponential with average and finite variance
\begin{equation*}
    \langle k \rangle_G = \frac{1+\beta}{\beta}, \qquad
    \Var_G(k) = \frac{1+\beta}{\beta^2}.
\end{equation*}
Notice that, though the dynamics of genes and genomes are mutually coupled and quantities such as the number of genomes grow at a rate that depends both on $\alpha$ and $\beta$, the  
%It is remarkable that, asymptotically in time, the 
two parameters of the model independently affect one or another degree distribution, asymptotically.
%though other quantities, such as the number of genomes, grows at a rate that depends both on $\alpha$ and $\beta$.  

\section{Modularity and correlations}

The analysis above shows that the combination of three simple mechanisms (HGT, functional innovation, and organismal innovation) is sufficient to generate the empirical degree distribution of genes and genomes. However, this analysis cannot determine whether the model induces significant correlations between pairs of nodes (for example, degree-degree correlations). Such correlations are characteristic of natural gene-genome bipartite networks: gene sharing is not equally likely among genomes, and genomes share genes above random expectations, resulting in the modularity observed in networks of DNA \cite{iranzo:2016a} and RNA viruses \cite{wolf:2018}. In terms of the network organization, modularity and correlations represent different ways of quantifying an underlying architectural feature—namely, the extent of overlap between the neighborhoods of genes or genomes—that measures deviations from a random assignment of links.

%In \cite{iranzo:2016a} a measure of clustering was applied to empirical gene-sharing networks, but that measure suffers from interpretative issues. After struggling with the possibility of generalizing clustering or overlap measures to bipartite networks, we left the issue aside because of its difficulties. Still, it might be interesting to include a measure that can reveal non-trivial correlations. 

In order to quantify the overlap between pairs of genes or pairs of genomes, we will use a measure introduced in \cite{johnson:2013} aimed at capturing the degree of nestedness in mono- and bipartite networks. The definition is as follows.   

Consider two empirical degree distributions, each corresponding to one of the node types in a bipartite network; in our case, these could be the distributions for genes and genomes obtained from our data sets or simulations of the mechanistic model with fixed parameters. In \cite{johnson:2013}, it is shown that the null model for a network with those distributions yields a value of overlap 
%(to fix ideas, consider a simulation of our model at time $t$),
\begin{equation*}
    \pi_{0} = \frac{N_g(t) \langle k_G^2 \rangle + 
    N_G(t) \langle k_g^2 \rangle}{\langle k_g \rangle \langle k_G \rangle (N_g(t)+N_G(t))^2},
\end{equation*}
using the nomenclature of our model. This value of overlap corresponds, therefore, to a configuration model\footnote{A configuration model is a family of random graph models designed to generate networks from a given degree sequence; that is, each node preserves its degree but the set of neighbors is randomly assigned.} \cite{bollobas:1980} with the empirically measured gene and genome degree distributions. 

The overlap has a local definition for each pair of nodes $\{i, j\}$ (regardless of whether they are of the same or different type),
\begin{equation*}
    \tilde{\pi}_{ij} = \frac{({\bf A}^2)_{ij}}{k_i k_j},
\end{equation*}
where ${\bf A}$ is the adjacency matrix of the bipartite graph. The square of the adjacency matrix reveals why this is a measure of overlap (or similarity) between nodes, because ${\bf A}^2$ is the number of walks of length two from node $i$ to node $j$. Considering that in bipartite networks there are no links between nodes of the same class (between pairs of genes or pairs of genomes), this is the number of genomes where a pair $\{i,j\}$ of genes is simultaneously present, or conversely, the number of genes shared by a pair $\{i,j\}$ of genomes. 

The network overlap is the average over all node pairs,
\begin{equation*}
    \tilde{\pi} = \frac{1}{N^2} \sum_{ij} \tilde{\pi}_{ij},
\end{equation*}
with $N=N_g(t)+N_G(t)$, 
and the relative overlap is finally defined as 
\begin{equation}
    \pi \equiv \frac{\tilde{\pi}}{\pi_{0}}.
    \label{eq:overlap}
\end{equation}

A notable property of this definition is that it corrects for heterogeneous degree distributions, which inherently induce positive nestedness (positive overlap) in finite networks \cite{johnson:2013}. Thus, if $\pi$ is significantly different from 1, there are degree-degree correlations beyond those caused by heterogeneous degree distributions. We measured this quantity for the analyzed data sets and for simulations of the model that best fit the data, and the resulting overlap values were compared (see below). 
%I believe this is a robust measure that speaks for degree-degree correlations without entering the clustering debate, but that could allow us to discuss limitations of the model (since it is to be expected that natural networks are more nested than simulated ones). However, I would expect deviations of simulated gene-sharing networks from the null model, due at least to the known correlation between node age and node degree in preferential attachment. 

\begin{figure}[tbp!]
    \centering
    \includegraphics[width=0.4\textwidth]{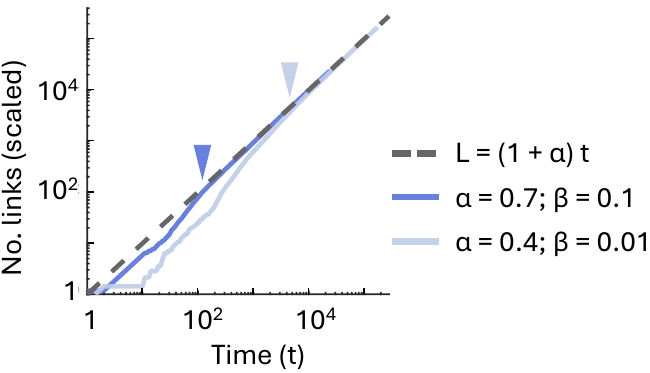}
    \caption[]{Evolution of the number of links in numerical simulations with $\epsilon=0$ (blue curves). The gray dashed line represents the asymptotic expectation in Eq. (\ref{eq:linklinear}). The transient before reaching the asymptotic behavior, approximately indicated with triangular markers, is significantly longer with smaller values of $\alpha$ and $\beta$, consistent with the fact that the initial condition is farther from the asymptotic average value of the degree distribution ($\langle k \rangle_g \simeq 3.5$ and $\langle k \rangle_G \simeq 100$ with $\alpha=0.4$ and $\beta=0.01$ vs $\langle k \rangle_g \simeq 2.4$ and $\langle k \rangle_G \simeq 11$ with $\alpha=0.7$ and $\beta=0.1$). To make all curves comparable, the number of links was scaled by a factor $(1+\alpha)^{-1}$.}
    \label{fig:evolution}
\end{figure}

%The main properties of this network are summarized in Table~\ref{tab:networks}. 
\begin{table*}[b!]
\begin{center}
\begin{tabular}{c|c|c|c|c|c|c|c|c}
\textbf{Data} & $N_g (\times 10^3)$ & $N_G (\times 10^3)$ & $L (\times 10^3)$ & $\langle k_g \rangle$ & $\langle k_G \rangle$ & $\alpha (\times 10^{-1})$ & $\beta (\times 10^{-2})$ & $\pi$ \\ 
\hline
dsDNA all & $3.4$ & $1.1$ & $98.6$ & $2.9$ & $91.8$ & $4.3$ $[2.1, 6.4]$ & $0.96$ $[0.7, 1.2]$ & $2.5$  \\
\hline
dsDNA core & $1.6$ & $1.1$ & $30.7$ & $19.5$ & $28.7$ & $-0.6$ $[-4.3, 2.7]$ & $3.3$ $[0.5, 5.6]$ & $2.7$ \\
\hline
RNA & $0.4$ & $2.9$ & $15.0$ & $36.2$ & $5.3$ & $-5.8$ $[-7.8, -3.9]$ & $28.1$ $[18.0, 38.0]$ & $2.2$ \\
\hline
Pangenomes & $42$ & $0.1$ & $136.7$ & $3.3$ & $1093$ & $3.3$ $[2.0, 4.5]$ & $0.08$ $[0.04, 0.12]$ & $1.0$ \\
%\hline
%dsDNA all & 33912 & 1074 & 98575 & 2.9 & 91.8 & 0.427 [0.209, 0.645] & 0.0096 [0.007, 0.012] & 2.5  \\
%\hline
%dsDNA core & 1576 & 1070 & 30692 & 19.5 & 28.7 & -0.062 [-0.428, 0.272] & 0.033 [0.005, 0.056] & 2.7 \\
%\hline
%RNA & 415 & 2856 & 15012 & 36.2 & 5.3 & -0.575 [-0.783, -0.390] & 0.281 [0.180, 0.380] & 2.2 \\
%\hline
%Pangenomes & 41520 & 125 & 136659 & 3.3 & 1093 & 0.326 [0.198, 0.454] & 0.0008 [0.0004, 0.0012] & 1.0 \\
\hline
\textbf{Model} & & & & & & & & \\
\hline
dsDNA all & $61$ $(1.5)$ & $1.5$ $(0)$ & $183.4$ $(4.7)$ & $3.0$ $(5)$ & $122$ $(3)$ & $4.8$ $[3.5, 7.2]$ & $0.81$ $[0.57, 0.93]$ & $0.80$ $(1)$ \\
\hline
dsDNA core & $7.4$ $(0.3)$ & $2.1$ $(0)$ & $63.0$ $(19.0)$ & $8.6$ $(1)$ & $29$ $(1)$ & $1.2$ $[0.7, 3.3]$ & $3.05$ $[0.76, 3.76]$ & $0.91$ $(1)$ \\
\hline
RNA & $1.0$ $(0)$ & $11.6$ $(0.5)$ & $51.4$ $(1.2)$ & $51$ $(1)$ & $4.4$ $(1)$ & $0.2$ $[0.2, 0.2]$ & $16.3$ $[5.7, 16.3]$ & $1.04$ $(1)$
%\hline
%dsDNA all & 60926 (1554) & 1500 (0) & 183449 (4652) & 3.000 (5) & 122 (3) & 0.48 [0.35, 0.72] & 0.0081 [0.0057, 0.0093] & 0.80 (1) \\
%\hline
%dsDNA core & 7359 (254) & 2140 (0) & 62982 (1894) & 8.6 (1) & 29 (1) & 0.12 [0.07, 0.33] & 0.0305 [0.0076, 0.0376] & 0.91 (1) \\
%\hline
%RNA & 1000 (0) & 11574 (477) & 51372 (1184)  & 51 (1) & 4.4 (1) & 0.02 [0.02, 0.02] & 0.163 [0.057, 0.163] & 1.04 (1) %\\ 
%\hline
%Pangenomes & $\infty$ & $\infty$ & $\infty$ & 3.0 & 988.5 & 0.326 & 0.0008 &  
\end{tabular}
\end{center}
\caption[]{Summary of variables and parameters obtained from data sets and the corresponding fits of the gene-sharing model. In the upper part of the table (\textbf{Data}) we show the size of empirical networks, the number of links $L$ they have, the average number of genomes where each gene family is found (i.e. gene average degree, $\langle k_g \rangle$) and the average number of genes per genome (genome average degree, $\langle k_G \rangle$). Parameters $\alpha$ and $\beta$ have been obtained by directly fitting the asymptotic degree distributions, Eqs. (\ref{eq:pkfinal}) and (\ref{eq:qkfinal}). Values of overlap $\pi$ have been obtained from Eq. (\ref{eq:overlap}) applied to real networks. In the lower part of the table (\textbf{Model}), we show the equivalent quantities averaged over realizations of the model with parameters that yield best fits to empirical degree distributions, following criteria to stop simulations as described in the main text. The value of $\pi$ has been obtained as an average over such realizations. Number in parentheses indicate the standard deviations. Numbers in brackets indicate the 95\% confidence intervals. See the main text for the application of the model to pangenomes.
%\textcolor{red}{Una vez se completen los datos podemos comentar cómo se compara la distribución asintótica con la numérica a tiempo finito. En el panel superior ajustamos los datos a una power-law (eso nos da un valor de $\alpha$) y a una exponencial para obtener $\beta$. En la parte inferior ponemos los parámetros que mejor ajustan simulando el modelo y haciendo mínimos cuadrados a las distribuciones; las que se obtienen de la simulación del modelo pueden estar lejos de la forma asintótica, y devolver por tanto valores distintos de $\alpha$ y $\beta$. Aunque en algunos casos no hay mayores diferencias (como en el pangenoma, donde de hecho no hay simulaciones para poner los parámetros de la parte inferior de la tabla), en otros casos sí existe, como para los virus de RNA. Jaime: Ok, creo que ya lo entiendo. Los valores de la parte superior se obtienen ajustando la distribución estacionaria. El único matiz es que esos valores no correspondían a las curvas ``estacionarias" de la Figura 3 original (en la última versión he modificado la figura para que corresponda con estos parámetros). En la parte inferior (Model) para el pangenoma he puesto $\infty$ en el número de genes, genomas y enlaces para subrayar que se trata de la solución asintótica. Si os parece inapropiado lo dejamos como ``--". Para la nestedness del modelo de pangenomas, ¿tenemos alguna aproximación analítica?}
}
\label{tab:networks}
\end{table*}

\section{Gene-sharing networks for viruses and prokaryotes}

We analyzed gene-sharing networks for three data sets: dsDNA viruses \cite{iranzo:2016a}, RNA viruses \cite{wolf:2018} and prokaryotic pangenomes, for which the bipartite network was reconstructed in this work. The characteristics of all networks are summarized in Table~\ref{tab:networks}.

In the case of dsDNA viruses, the two classes of nodes correspond to orthologous gene families and viral genomes. The dataset included all families of dsDNA viruses known at the time of our previous analysis \cite{iranzo:2016a} and some related mobile genetic elements. Beyond an emerging modular structure that largely reflected accepted taxa, it was shown that the degree distribution for gene families was compatible with a power-law with an exponent close to $-2$, while that of viral genomes was much flatter and considered to be roughly uniform. Two different bipartite networks were generated from this dataset: (i) a network consisting of all orthologous gene families (including singletons and highly divergent genes), denoted `dsDNA all', and (ii) a core network in which only genes with a loss rate below unity were retained, denoted `dsDNA core' (see \cite{iranzo:2016a} for details). 
%The characteristics of the two networks are summarized in Table~\ref{tab:networks}. 

RNA virus gene sharing networks were constructed in \cite{wolf:2018}. In this case, gene nodes correspond to protein domains identified from a representative ensemble of RNA virus genomes (see Materials and Methods)
%, including representative members of ICTV-approved virus families and unclassified virus groups.
%---that was annotated and manually curated. Network details are reported in Table~\ref{tab:networks}. In this case, as well 
As for the dsDNA core network, 
%gene-sharing network, 
the data was filtered in order to retain only evolutionarily informative genes, as in the original publications. 
%Network details are reported in Table~\ref{tab:networks}.

Finally, we analyzed a network reconstructed from the pangenomes of 125 bacterial and archaeal species analyzed in \cite{manzano:2023}. The pangenome network was built by connecting each gene family to all the pangenomes in which it is present (see Materials and Methods). 

\section{Comparison of gene and genome degree distributions for model-predicted, simulated and empirical gene-sharing networks}

An assumption of our mean-field approximation was that the number of links in a network grows linearly with time \eqref{eq:linklinear}. Analytical results indicate that this assumption is self-consistent, and simulations confirm an asymptotically linear increase in the number of links (Figure~\ref{fig:evolution}). These simulations further reveal a transient regime, the duration of which depends on the model parameters. The variable length of these transients suggests that natural systems might lie at different distances from the long-time limit, depending on their initial conditions and the parameters that govern their dynamics.

To explore the regions of the $\{\alpha, \beta\}$ plane where the degree distributions fit well the model predictions, we conducted numerical simulations of the model across a broad range of parameter values and compared the resulting degree distributions with empirical ones. The optimal values of $\alpha$ and $\beta$ were determined by minimizing the overall least-squares error between simulated and empirical degree distributions (see Materials and Methods). 

\begin{figure*}[tbp!]
\includegraphics[width=\textwidth]{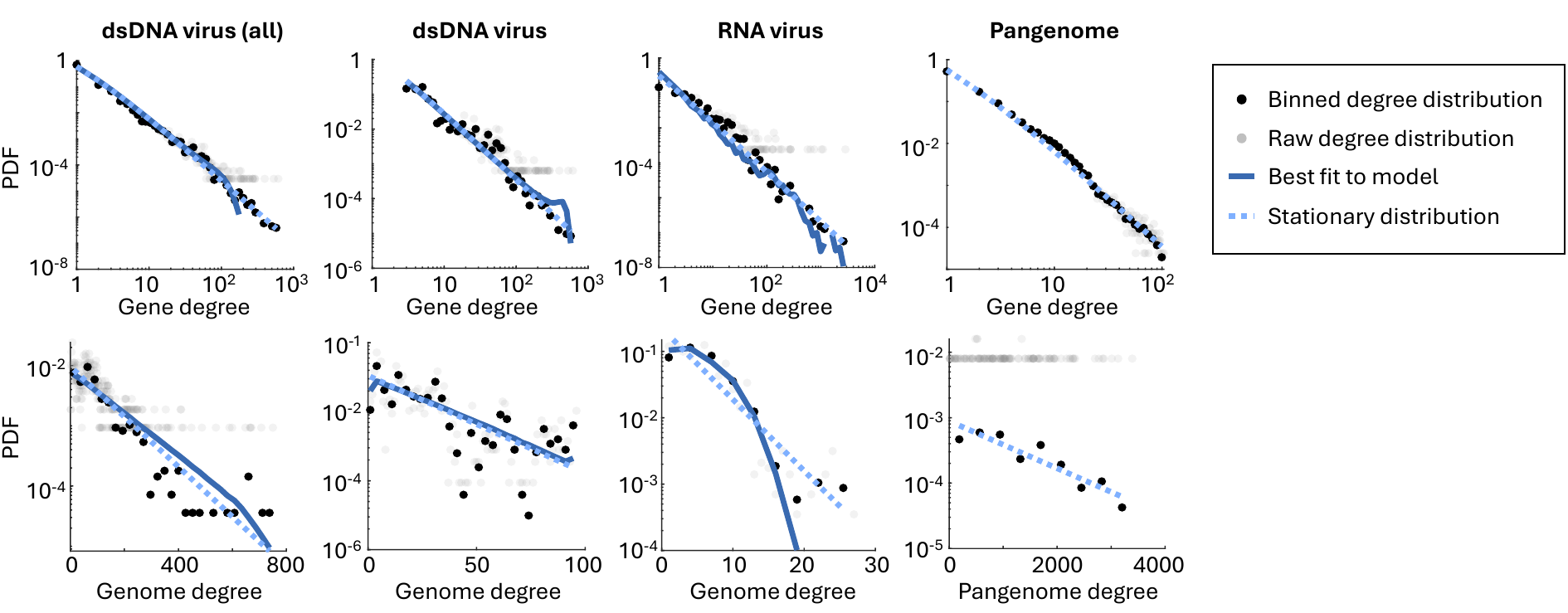}
\caption[]{Degree distributions of real-world gene-sharing networks and fits to the model. 
%The label ``dsDNA virus (all)'' corresponds to the complete dsDNA virus network, which includes singletons (genes with no known homologs) and highly divergent genes. The two other viral networks (dsDNA and RNA) were filtered to include only evolutionarily informative genes, as in the original publications. 
Best fits were obtained by minimizing the total squared deviation between the binned empirical distributions and Monte Carlo simulations of the model, except in the case of the pangenome network, that was directly fitted using the analytical expression for the stationary distribution. Fitting parameters are given in Table~\ref{tab:networks}. 
%Fitted parameters: dsDNA virus (all) $\alpha = 0.48$, $\beta = 0.0081$; dsDNA virus $\alpha = 0.12$, $\beta = 0.0305$; RNA virus $\alpha = 0.02$, $\beta = 0.163$; pangenome $\alpha = 0.326$, $\beta = 0.0011$.
%\textcolor{red}{I modified the figure so that the dotted line represents the best stationary fit (section ``Data'' in Table 1), not the stationary distribution given the parameters of the (non-stationary) simulation-based fit. In case you prefer the older figure, it is now named Fig\_dd\_fits.old.pdf}
}
\label{fig:DegDistrFit}
\end{figure*}

Figure~\ref{fig:DegDistrFit} summarizes the results of fitting empirical data with simulation results for the four networks. All probability distributions show a good agreement with numerical results for a pair of $\{\alpha, \beta\}$ values (see Table~\ref{tab:networks}). For comparison, we also show fits to the degree distributions used by fitting their asymptotic forms, Eqs.~(\ref{eq:pkfinal}) and (\ref{eq:qkfinal}); these values are given in the upper part of Table \ref{tab:networks}. Gene degree distributions are in all cases remarkably close to their asymptotic shape, although some deviations are apparent for the smaller, noisier networks (especially those for RNA viruses). Genome degree distributions are also noisier because of their smaller sizes; however, deviations from the expected asymptotic exponential shape only stand out for the RNA virus network. 

\begin{figure}[tbp!]
\begin{center}
\includegraphics[width=0.48\textwidth]{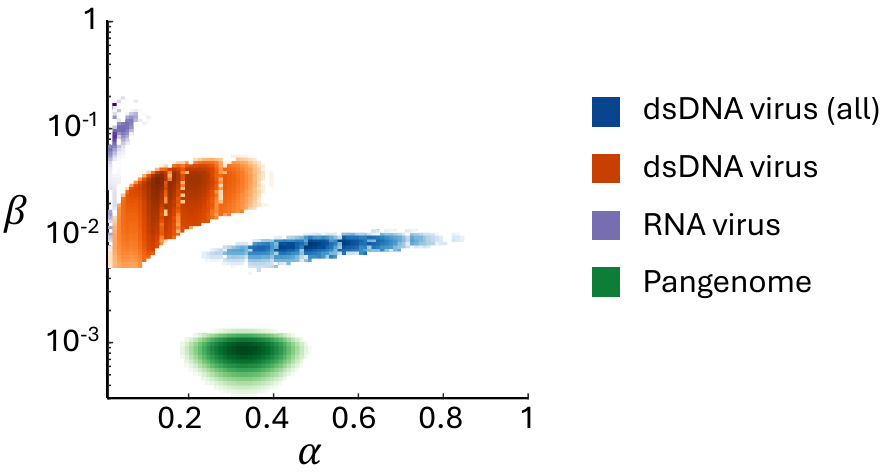}
\caption[]{Regions of the parameter space that provide a good fit to the empirical networks (total squared deviation less than twice the minimum value). The regions of parameters that fit each empirical gene-sharing network are non-overlapping, hinting at dynamical processes occurring at significantly different rates for the systems analyzed.}
\label{fig:FitRegions}
\end{center}
\end{figure}

The model parameter values do not need to be fine-tuned to fit empirical data. Indeed, relatively broad regions of the parameter space yield nearly optimal fits to the data (Fig.~\ref{fig:FitRegions}). Notably, however, these regions do not overlap for different datasets, suggesting that the processes of gene and genome innovation run at substantially different rates in each case (see Discussion). 

Table \ref{tab:networks} compares the results of fitting data to the asymptotic degree distributions (upper section) with the model parameters that provide the best fit to the empirical data (lower section). Remarkably, the values of $\alpha$ and $\beta$ obtained through these two approaches are closely similar. The negative values of $\alpha$ observed for the dsDNA core and RNA networks warrant discussion. By construction, the model does not permit negative $\alpha$ values, yielding instead small positive ones in both cases. However, nothing prevents negative values when \eqref{eq:pkfinal} is fitted directly to the data. It is also worth noting that the RNA network exhibits a systematic downward curvature, indicating that fits over different regions would produce distinct exponent values. Simulations better capture such deviations from linearity, which reflect the transient dynamics preceding the asymptotic regime. Overall, the model—with only two parameters— captures the central trends in the data and explains \textgreater95\% of the variance of the gene degree distributions and 45-90\% of  the variance of the genome degree distributions.

It is noteworthy, however, that the simulated networks were two to three times larger than the empirical ones. We believe this discrepancy arises because the former include all nodes and links generated by the model, whereas empirical networks represent only a partial sampling of a much larger pool of genomes (in other words, the genomes included in the dsDNA and RNA virus networks only cover a limited, and actually, small fraction of all extant viruses, and the same applies to pangenomes). To test this explanation, we repeated the measurements using subsamples of the simulated networks with sizes comparable to those of the empirical data. We found that the degree distributions of the subsampled networks did not differ substantially from those of the complete networks (Fig.~S1).
%Si esta figura va en material suplementario, hay que nombrarla a mano. Habrá que hacer ese fichero, quizá incluyendo las simulaciones nuevas con especiación.
Thus, we can conclude that partial networks obtained from a random subset of genomes are informative about the structure of the complete network.

The values of the overlap $\pi$ are also informative. As expected, the model yields values close to 1, consistent with the lack of correlations in link assignment inherent to the model design. In contrast, the dsDNA all, dsDNA core, and RNA networks display significantly higher values, reflecting the clustering previously identified in these networks. The pangenome network, however, shows random overlap among nodes. Because only one species per genus was included in this network, we interpret such lack of correlations as a consequence of massive turnover of accessory genes throughout the evolution of bacterial genera, leading to gene sharing patterns that are compatible with random assignment of links.

\begin{figure}[tbp!]
\begin{center}
    \includegraphics[width=0.48\textwidth]{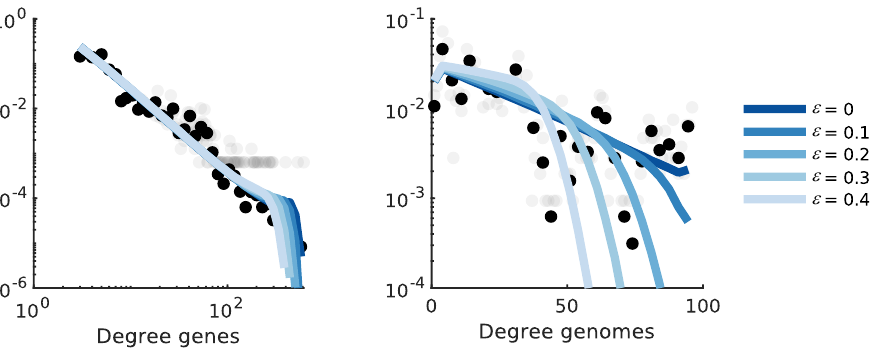}
\caption[]{Effect of gene loss on the gene and genome degree distributions. While gene degree distributions are barely affected by gene loss, genome degree distributions shorten their range and the fit to data worsens. Data correspond to the core dsDNA virus network.}
\label{fig:Deletion}
\end{center}
\end{figure}

All the above results were obtained in the limit of no gene loss, $\epsilon=0$. Although degree distributions were well fit in this limit, the resulting networks contained many more links than the empirically observed networks (Table~\ref{tab:networks}). As discussed above, this discrepancy can be attributed to the implicit sampling of genomes in empirical networks and disappears if the simulated networks are equally sampled. Analytical derivations of network growth and degree distributions with nonzero gene loss ($\epsilon>0$) are not feasible; therefore, we resorted to simulations to investigate the effect of gene loss. We did not observe significant differences in the fits even if $\alpha$ and $\beta$ were fixed (Fig.~\ref{fig:Deletion}). Gene degree distributions were minimally affected, which is to be expected considering that these distributions emerge from a multiplicative process (preferential attachment) whereas gene loss is an additive process. In genome degree distributions, the shape of their rightmost tail changed, showing faster decay the larger the value of $\epsilon$, deviating from observations. Thus, the rate of gene loss apparently should take values $\epsilon \lesssim 0.1$ 
%close to zero 
to fit real data. %\textcolor{red}{The number of links...} 
%it cannot be discarded that optimized fits varying the three parameters simultaneously could return fits as good as with $\epsilon=0$ for degree distributions and also match the total number of links in the network

\section{Discussion}

We propose here that the architecture of bipartite gene-sharing networks is rooted in two dynamical processes: %preferential attachment of genes 
HGT of genes proportional to their abundance in the genome ensemble and innovation, be it due to the capture of new genes or to the emergence of new minimal genomes. This conjecture is supported by analytical and numerical results obtained with our generative model for gene sharing networks that, with only two free parameters---probability of a new gene capture ($\alpha$) and probability of a new genome emergence ($\beta$)---successfully recovers the observed degree distributions of genes and genomes. We derived analytical expressions for these distributions evolving in the long term and show, by comparison with numerical simulations, that despite the transient nature and limited size of the empirical networks, they can be well approximated by the asymptotic forms of the degree distribution. The analytical expressions for the node degree distributions were obtained under the no gene loss assumption. However, numerical investigation of the effect of gene loss showed that this process had a limited effect on the main statistical properties of gene sharing networks even at relatively high loss rates. Nevertheless, to obtain degree distributions that qualitatively agreed with those of the empirical networks, gene loss rates must remain lower than gain rates. This finding suggests a fundamental distinction between the evolution of prokaryotic genomes, which is driven, in large part, by the intrinsic bias towards gene loss that can be counter-balanced by gene gain under selection for functional diversification \cite{snel:2002, koonin:2008, puigbo:2014, sela:2016, iranzo:2017b, bobay:2018}, %PMID: 30314447
and the evolution of prokaryotic pangenomes and viral genomes, where gene gain appears to be the dominant process. This conclusion is compatible both with the vast size of most of the well-studied prokaryotic pangenomes, many of which are open (that is, show no signs of saturation with the addition of new genomes) \cite{puigbo:2014, horesh:2021, dewar:2024}, and evolutionary reconstructions for genomes of large DNA viruses \cite{yutin:2014, koonin:2019}.  %, though it could be used to adjust the number of links in the network, \textcolor{red}{which in the model is systematically larger than in natural networks -- Due to sampling, or something else?}.

 The parameters $\alpha$ and $\beta$ can be interpreted in the context of the data analyzed. The largest values of $\alpha$, which represents the rate of introduction of new genes into the system, correspond to the all-gene network for dsDNA viruses, followed by the pangenome network, the dsDNA core network and, finally, the RNA virus network. The differences in the $\alpha$ values likely reflects the permissiveness of genomes and pangenomes to accept new genes, which correlates with their size, but also the trend for capture and exaptation of host genes that appears to dominate the evolution of large DNA viruses \cite{koonin:2022}. %PMID: 35834963
 The differences in $\alpha$ could also be affected by the gene inclusion criteria applied to each network. Specifically, the emergence or capture of gene families that eventually become central (that is, part of the RNA and dsDNA virus core networks) is substantially less frequent than the \emph{de novo} emergence or acquisition of less central, accessory genes (which are included in the all-gene dsDNA virus network) \cite{wolf:2016}.  %PMID: 27819663. 
Conversely, in its turn, the largest value of $\beta$ was obtained for to RNA viruses, followed by dsDNA viruses (considering either all genes or only core genes) and, finally, the pangenomes. Indeed, the rate of organismal innovation is likely to be the highest for RNA viruses, with their small genomes, where acquisition of a single gene has the potential to found a new group of viruses \cite{wolf:2018}, %PMID: 30482837, 
and the lowest for prokaryotic pangenomes, consisting of many thousands of genes

%\textcolor{red}{Are there other known processes that qualitatively agree with the rates begin larger/smaller in different systems? Are the relative rates consistent with known evolutionary characteristics of the different groups? My feeling is that the differences in $\alpha$ are primarily driven by the criteria used to filter genes (hypothetical genes in dsDNA virus full, conserved gene families in pangenomes, and evolutionarily retained genes in dsDNA virus core and RNA virus networks). To a lesser extent, $\alpha$ could be related to the permissiveness of genomes and pangenomes to accept new genes, which correlates with their size.}

%\subsection{Yule-Simon distribution}

%A time step in our simulations corresponds to an HGT attempt. The occurrence of such events is not homogeneous in time, since HGT is less frequent in small systems, its likelihood increasing as the system grows. Actually, in real time HGT events take place at an exponentially growing pace, since the total number of genes grows exponentially in real time \cite{manrubia:2002}. 

The relative overlap in simulated networks is very close to unity, indicating absence of correlations, whereas overlap in viral gene-sharing networks is significantly higher. There are various processes that could cause correlations but were not considered in our model, such as concerted evolution of functionally linked genes, variations in the selection strength experienced by different gene categories, or heterogeneous HGT rates across different groups of genomes. Environmental pressure, whereby genes that are important for adaptation to a specific niche are convergently acquired by viruses that occupy that niche (for example, specialized lysins in mycophages \cite{catalao:2018}), %PMID: 30110929
could also contribute to nonrandom overlap in gene content. Arguably, however, the main cause of correlations in real gene-sharing networks is phylogenetic relatedness among genomes. In our model, each genome results from an independent line of descent, with relatedness due only to HGT. This, obviously oversimplified, approach could well approximate the relationships between distantly related organisms or viruses, considering the small size of evolutionarily stable gene cores, but breaks down for more closely related genomes. 
%Straightforward extensions of the model that generate tree-like phylogenies, for example by decoupling speciation from \emph{de novo} genome seeding, generally fail to reproduce the degree distributions observed in nature. 
Reconciling the simple design of this model with tree-like evolution, while preserving the basic statistical properties of gene-sharing networks, remains an open task for future investigation.

\section{Materials and methods}

\subsection{Probability of link formation}

We can extract information from the master \eqref{eq:master} by introducing the generating function
\begin{equation*}
    g(z,t)\equiv\sum_{k\ge 1}g_k(t)z^k.
\end{equation*}
%Then, multiplying \eqref{eq:master} by $z^k$, summing over $k\ge 1$, and using the identities
%\begin{align*}
%    &\sum_{k\ge 1}z^k\mathbb{E}^{-1}f(k)=z\sum_{k\ge 0}z^kf(k), \\
%    &\sum_{k\ge 1}kz^kf(k)=z\frac{\partial}{\partial z}\sum_{k\ge 0}z^kf(k),
%\end{align*}
%equation~\eqref{eq:master} becomes
Differentiating $g(z,t)$ with respect to $t$ and substituting \eqref{eq:master} we
obtain
\begin{equation}
    g_t=\frac{(1-z)z}{L(t)}\left[\frac{1-\beta}{N_G(t)}zg_z-g\right]_z+\alpha z.
    \label{eq:mastergenerating}
\end{equation}
(Subscripts denote partial derivatives.) Notice that $g(1,t)=N_g(t)$, so setting $z=1$ in \eqref{eq:mastergenerating} yields the known equation $\dot N_g=\alpha$. On the other hand, $g_z(1,t)=L(t)$. If we differentiate \eqref{eq:mastergenerating} with respect to $z$,
%we obtain
%\begin{align*}
%    g_{t,z}=&\, \frac{1-2z}{L(t)}\left[\frac{1-\beta}{N_G(t)}(g_z+zg_{zz})-g_z\right] \\
%    &+\frac{(1-z)z}{L(t)}\left[\frac{1-\beta}{N_G(t)}zg_z-g\right]_{zz}+\alpha.
%\end{align*}
%Setting
set $z=1$, use the identity $g_z(1,t)+g_{zz}(1,t)=\sum_{k\ge 1}k^2g_k(t)$, and compare with \eqref{eq:linkgrowth}, we finally obtain \eqref{eq:Pf}.

\subsection{Asymptotic degree distribution of genes}

As $t\to\infty$, $g(z,t)\sim N_g(t)p(z)$, with $p(z)$ the generating function of $p_k$. Substituting in \eqref{eq:mastergenerating} and taking into account that in this limit
\begin{equation*}
    \dot N_g=\alpha, \qquad \frac{N_g(t)}{L(t)}\to\frac{\alpha}{L_{\infty}}, \qquad \frac{1-\beta}{N_G(t)}\to 0,
\end{equation*}
where $L_{\infty}\equiv\alpha+\beta+(1-\beta)P_{\infty}$, we find that $p(z)$ satisfies the equation
\begin{equation*}
    p'+\frac{L_{\infty}}{z(1-z)}p=\frac{L_{\infty}}{1-z}, \qquad p(0)=0,
\end{equation*}
whose solution is
\begin{equation*}
    p(z)=L_{\infty}\left(\frac{1-z}{z}\right)^{L_{\infty}}
    \int_0^z\frac{u^{L_{\infty}}}{(1-u)^{L_{\infty}+1}}\,du.
\end{equation*}
With the change of variable $u=z(1-x)/(1-zx)$ this integral becomes
\begin{equation}
    p(z)=L_{\infty}z\int_0^1\frac{(1-x)^{L_{\infty}}}{1-zx}\,dx.
    \label{eq:psfinal}
\end{equation}

Setting $z=1$ in this expression immediately yields $p(1)=1$, as expected. On the other hand, from the same expression $p'(1)=L_{\infty}/(L_{\infty}-1)$. As  $g_z(1,t)\sim L_{\infty}t$ and, moreover, $g_z(1,t)\sim \alpha tp'(1)$, it follows that $L_{\infty}=1+\alpha$ (equivalently, $P_{\infty}=1$, consistent with the direct calculation from \eqref{eq:Pf}).
%In other words, asymptotically in time (which implies a large enough system), the link between a gene chosen with a probability proportional to its degree and a randomly chosen genome does not exist with probability $P_{\infty}=1$.  

We can now expand the integrand in \eqref{eq:psfinal}
\begin{align*}
     p(z) &=(1+\alpha)\sum_{k=1}^{\infty}z^k
     \int_0^1x^{k-1}(1-x)^{1+\alpha}\,dx \\
     &=(1+\alpha)\sum_{k=1}^{\infty}z^k
     \frac{\Gamma(k)\Gamma(2+\alpha)}{\Gamma(k+2+\alpha)},
\end{align*}
from which we readily obtain \eqref{eq:pkfinal}.

\subsection{Construction of empirical gene-sharing networks}

RNA virus gene-sharing networks were constructed as described in \cite{wolf:2018}. Gene nodes correspond to a set of protein domains identified using a representative ensemble of RNA virus genomes---including representative members of ICTV-approved virus families and unclassified virus groups---that was annotated and manually curated. 

The pangenome network was reconstructed from the pangenomes of 125 bacterial and archaeal species analyzed in \cite{manzano:2023}. The Genome Taxonomy Database (GTDB) was parsed to identify species with at least 15 available high-quality genomes, selecting at most one species per genus and discarding all medium and low-quality genomes. For computational reasons, species with more than 100 high-quality genomes were subsampled to keep a representative set of 100 genomes. Gene families were defined by mapping predicted open reading frames to the eggNOG database at the root level. Core genes (those present in all the genomes of a species) were not included in the pangenome network because these genes are known to experience strong selection against horizontal gene transfer, which produces trends incompatible with neutral gain and loss models \cite{lobkovsky:2013}. The pangenome network was built by connecting each gene family to all the pangenomes in which it is present. 

\subsection{Numerical simulations and parameter fitting}

To explore the regions of the $\{\alpha,\beta\}$ parameter space compatible with empirical data, we performed a comprehensive parameter sweep over $\alpha$ (linearly spaced from 0.01 to 1 in 99 steps) and $\beta$ (logarithmically spaced from 0.001 to 1 in 99 steps), with $\epsilon$ fixed at zero (no gene loss). For each parameter pair, we ran 1000 independent simulations.

Each simulation proceeded until a time $t_s$ when it satisfied both a minimum number of genes and genomes with heuristic thresholds tailored to each virus group:

\begin{enumerate}\setlength{\itemsep}{-2pt}
    \item dsDNA virus (full): $N_g(t_s)>50\,000$ and $N_G(t_s)>1\,500$; 
    \item dsDNA virus (core): $N_g(t_s)>3\,200$ and $N_G(t_s)>2\,140$; 
    \item RNA virus: $N_g(t_s)>1\,000$ and $N_G(t_s)>5\,000$; 
    \item Pangenomes: suitable fits to distributions required increasing the number of genes and genomes to sizes that demanded prohibitively large computational time. We therefore fitted the long-time solution of the model, Eqs.~(\ref{eq:pkfinal}) and (\ref{eq:qkfinal}).
\end{enumerate}

We also imposed maximum values due to computational limits: dsDNA virus (full): 80000 genes, 4000 genomes; dsDNA virus (core): 30000 genes, 5000 genomes; RNA: 1500 genes, 18000 genomes in order to have a reasonable memory usage given the network properties. Runs that attained the maximum value for one of the variables without having reached the minimum required for the other one were discarded. 
%\textcolor{red}{¿Cuál es el valor máximo permitido? Adapted to obtain a reasonable memory usage given the network properties. dsDNA virus (full): 80000 genes, 4000 genomes; dsDNA virus (core): 30000 genes, 5000 genomes; RNA: 3000 genes, 10000 genomes.} 
Attempting a weaker rule (stopping when just one threshold was reached) failed to reproduce the heavy tails found in real data because there was not enough time for the simulations to generate the extreme-degree nodes (in genes or genomes) observed in empirical networks.

Once simulations finished, simulated degree distributions were compared to empirical ones via least-squares fitting. Gene degrees were binned logarithmically and genome degrees were binned linearly ($50$ bins for genes and $30$ bins for genomes in dsDNA networks; $50$ bins for genes and $10$ bins for genome in RNA networks). Bins with no data were omitted. The optimal $\alpha$ and $\beta$ were those that minimized the overall error (lower part of Table \ref{tab:networks}). 

%\section{Data availability} 
%The table of RNA virus genomes and their encoded protein domains (the RNA virus gene-sharing network) can be downloaded from the original publication (Data Set S5) \cite{wolf:2018} and from ftp://ftp.ncbi.nlm.nih.gov/pub/wolf/\_suppl/rnavir18/RNAvirome\_DatasetS5.xls. The gene-sharing networks of dsDNA viruses and pangenomes are deposited in Zenodo (DOI:10.5281/zenodo.19445652 and 10.5281/zenodo.19441269, respectively). 
%The model was simulated using compiled C++ scripts. The source code for the simulations is available as a Supplementary File.

\section{Acknowledgments}

This work was funded by MICIU/AEI/10.13039/501100011033 and by ERDF/EU (PGE) ``A way of making Europe'', through grants PID2023-147963NB-C21, PID2022-141802NB-I00, PID2019-106618GA-I00, and CNS2023-145430.
E.V.K. is supported through the intramural program of the U.S. National Institutes of Health.

%\section{Author contributions}

%SM and JAC designed research; JI, PJ, SM, and JAC performed research; JI and PJ analyzed data; all authors discussed and interpreted results; JI, EVK, SM, and JAC wrote the paper.

%\section{Competing interests}

%The authors declare no competing interests.

%\bibliographystyle{apsrev4-2}
\bibliographystyle{apsrev4-2}
\bibliography{references}

@STRING{PNAS = {Proc. Natl. Acad. Sci. USA}}

@STRING{PTRSB = {Phil. Trans. R. Soc. Lond. B}}

@STRING{TREE = {Trends Ecol. Evol.}}

@article{barabasi:1999,
    author = "A. L. Barab\'asi and R. Albert",
    title = "Emergence of Scaling in Random Networks",
    year = 1999,
    journal = "Science",
    volume = 286,
    pages = "509--512"
}

@article{basten:1991,
    author = "C. J. Basten and M. E. Moody",
    title = "A branching-process model for the evolution of transposable elements incorporating selection",
    year = 1991,
    journal = "J. Math. Biol.",
    volume = 29,
    pages = "743--761"
}

@article{bobay:2018,
    author = {Bobay, L. M. and Ochman, H.},
    title = {Factors driving effective population size and pan-genome evolution in bacteria},
    journal = {BMC Evol. Biol.},
    volume = {18},
    pages = {153},
    year = {2018}
}

@article{bollobas:1980,
title = {A Probabilistic Proof of an Asymptotic Formula for the Number of Labelled Regular Graphs},
journal = {European Journal of Combinatorics},
volume = {1},
number = {4},
pages = {311-316},
year = {1980},
issn = {0195-6698},
doi = {https://doi.org/10.1016/S0195-6698(80)80030-8},
url = {https://www.sciencedirect.com/science/article/pii/S0195669880800308},
author = {B{\'e}la Bollob{\'a}s}
}

@article {booth:2016,
	Title = {The Modern Synthesis in the Light of Microbial Genomics},
	Author = {Booth, Austin and Mariscal, Carlos and Doolittle, W Ford},
	DOI = {10.1146/annurev-micro-102215-095456},
	Volume = {70},
	Month = {September},
	Year = {2016},
	Journal = {Ann. Rev. Microbiol.},
	ISSN = {0066-4227},
	Pages = {279--297},
	URL = {https://doi.org/10.1146/annurev-micro-102215-095456},
}

@Article{catalao:2018,
author = {Catal{\~a}o, Maria Jo{\~a}o and Pimentel, Madalena},
title = {Mycobacteriophage Lysis Enzymes: Targeting the Mycobacterial Cell Envelope},
journal = {Viruses},
volume = {10},
year = {2018},
number = {8},
article-number = {428},
URL = {https://www.mdpi.com/1999-4915/10/8/428},
PubMedID = {30110929},
ISSN = {1999-4915},
DOI = {10.3390/v10080428}
}

@article{charlesworth:1983,
    author = "B. Charlesworth and D. Charlesworth",
    title = "The population dynamics of transposable elements",
    year = 1983,
    journal = "Genet. Res. Camb.",
    volume = 42,
    pages = "1--27"
}

@article{corel:2018,
    author = "E. Corel and R. M\'eheust and A. K. Watson and J. O. McInerney and P. Lopez and E. Bapteste",
    title = "Emergence of Scaling in Random Networks",
    year = 2018,
    journal = "Mol. Biol. Evol.",
    volume = 35,
    pages = "899--913"
}

@article{dewar:2024,
author = {Anna E. Dewar  and Chunhui Hao  and Laurence J. Belcher  and Melanie Ghoul  and Stuart A. West },
title = {Bacterial lifestyle shapes pangenomes},
journal = PNAS,
volume = {121},
number = {21},
pages = {e2320170121},
year = {2024},
doi = {10.1073/pnas.2320170121},
URL = {https://www.pnas.org/doi/abs/10.1073/pnas.2320170121},
eprint = {https://www.pnas.org/doi/pdf/10.1073/pnas.2320170121}
}

@article{doolittle:1999,
author = {W. Ford Doolittle},
title = {Phylogenetic Classification and the Universal Tree},
journal = {Science},
volume = {284},
number = {5423},
pages = {2124-2128},
year = {1999},
doi = {10.1126/science.284.5423.2124},
URL = {https://www.science.org/doi/abs/10.1126/science.284.5423.2124},
eprint = {https://www.science.org/doi/pdf/10.1126/science.284.5423.2124}
}

@article{horesh:2021,
   author = "Horesh, Gal and Taylor-Brown, Alyce and McGimpsey, Stephanie and Lassalle, Florent and Corander, Jukka and Heinz, Eva and Thomson, Nicholas R.",
   title = "Different evolutionary trends form the twilight zone of the bacterial pan-genome", 
   journal= "Microbial Genomics",
   year = "2021",
   volume = "7",
   number = "9",
   pages = "",
   doi = "https://doi.org/10.1099/mgen.0.000670",
   url = "https://www.microbiologyresearch.org/content/journal/mgen/10.1099/mgen.0.000670",
   publisher = "Microbiology Society",
   issn = "2057-5858"
  }

@article{hugues:2008,
title = {The power-law distribution of gene family size is driven by the pseudogenisation rate's heterogeneity between gene families},
journal = {Gene},
volume = {414},
number = {1},
pages = {85--94},
year = {2008},
issn = {0378-1119},
doi = {https://doi.org/10.1016/j.gene.2008.02.014},
url = {https://www.sciencedirect.com/science/article/pii/S0378111908000802},
author = {Timothy Hughes and David A. Liberles}
}

@article{huynen:1998,
    author = {Huynen, M A and van Nimwegen, E},
    title = {The frequency distribution of gene family sizes in complete genomes.},
    journal = {Molecular Biology and Evolution},
    volume = {15},
    number = {5},
    pages = {583--589},
    year = {1998},
    month = {05},
    issn = {0737-4038},
    doi = {10.1093/oxfordjournals.molbev.a025959},
}

@article{iranzo:2016a,
    author = "J. Iranzo and M. Krupovic and E. V. Koonin",
    title = "The Double-Stranded {DNA} Virosphere as a Modular Hierarchical Network of Gene Sharing",
    year = 2016,
    journal = "mBio",
    volume = 7,
    pages = "e00978"
}

@article{iranzo:2016b,
    author = "J. Iranzo and E. V. Koonin and David Prangishvili and M. Krupovic",
    title = "Bipartite Network Analysis of the Archaeal Virosphere: Evolutionary Connections between Viruses and Capsidless Mobile Elements",
    year = 2016,
    journal = "J. Virol.",
    volume = 90,
    pages = "11043--11055"
}

@article{iranzo:2016c,
    author = "J. Iranzo and P. Puigb{\`o} and A. E. Lobkovsky and Y. I. Wolf and E. V. Koonin",
    title = "Inevitability of Genetic Parasites",
    year = 2016,
    journal = "Genome Biol. Evol.",
    volume = 8,
    pages = "2856--2869"
}

@article{iranzo:2017a,
    author = "J. Iranzo and M. Krupovic and E. V. Koonin",
    title = "A network perspective on the virus world",
    year = 2017,
    journal = "Commun. Integr. Biol.",
    volume = 10,
    pages = "e1296614"
}

@article{iranzo:2017b,
    author = "J. Iranzo and J. A. Cuesta and S. Manrubia and M. I. Katsnelson and E. V. Koonin",
    title = "Disentangling the effects of selection and loss bias on gene dynamics",
    year = 2017,
    journal = PNAS,
    volume = 114,
    pages = "E5616--E5624"
}

@article{jang:2019,
    author = "H. B. Jang and B. Bolduc and O. Zablocki and J. H. Kuhn and S. Roux and E. M. Adriaenssens and J. R. Brister and A. M. Kropinski and M. Krupovic and R. Lavigne and D. Turner and M. B. Sullivan",
    title = "Taxonomic assignment of uncultivated prokaryotic virus genomes is enabled by gene-sharing networks",
    year = 2019,
    journal = "Nat. Biotechnol.",
    volume = 37,
    pages = "632--639"
}

@article{johnson:2013,
    doi = {10.1371/journal.pone.0074025},
    author = {Jonhson, Samuel AND Dom{\'\i}nguez-Garc{\'\i}a, Virginia AND Mu{\~n}oz, Miguel A.},
    journal = {{PLoS ONE}},
    publisher = {Public Library of Science},
    title = {Factors Determining Nestedness in Complex Networks},
    year = {2013},
    month = {09},
    volume = {8},
    url = {https://doi.org/10.1371/journal.pone.0074025},
    pages = {1-7},
    number = {9}
}

@article{karev:2002,
    author = "G. P. Karev and Y. I. Wolf and A. Y. Rzhetsky and F. S. Berezovskaya and E. V. Koonin",
    title = "Birth and death of protein domains: a simple model of evolution explains power law behavior",
    year = 2002,
    journal = "BMC Evol. Biol.",
    volume = 2,
    pages = "18"
}

@article{karev:2004,
    author = "G. P. Karev and Y. I. Wolf and F. S. Berezovskaya and E. V. Koonin",
    title = "Gene family evolution: an in-depth theoretical and simulation analysis of non-linear birth-death-innovation models",
    year = 2004,
    journal = "BMC Evol. Biol.",
    volume = 4,
    pages = "32"
}

@article{koonin:2002,
    author = "E. V. Koonin and Y. I. Wolf and G. P. Karev",
    title = "The structure of the protein universe and genome evolution",
    year = 2002,
    journal = "Nature",
    volume = 420,
    pages = "218--223"
}

@article{koonin:2008,
    author = {Koonin, Eugene V. and Wolf, Yuri I.},
    title = {Genomics of bacteria and archaea: the emerging dynamic view of the prokaryotic world},
    journal = {Nucleic Acids Res.},
    volume = {36},
    number = {21},
    pages = {6688-6719},
    year = {2008},
    month = {10},
    issn = {0305-1048},
    doi = {10.1093/nar/gkn668}
}

@incollection{koonin:2019,
title = {{Chapter Five - Evolution of the Large Nucleocytoplasmic DNA Viruses of Eukaryotes and Convergent Origins of Viral Gigantism}},
editor = {Margaret Kielian and Thomas C. Mettenleiter and Marilyn J. Roossinck},
booktitle = {Advances in Virus Research},
publisher = {Academic Press},
volume = {103},
pages = {167-202},
year = {2019},
issn = {0065-3527},
doi = {https://doi.org/10.1016/bs.aivir.2018.09.002},
url = {https://www.sciencedirect.com/science/article/pii/S0065352718300551},
author = {Eugene V. Koonin and Natalya Yutin}
}

@article{koonin:2022,
title = {The logic of virus evolution},
journal = {Cell Host Microbe},
volume = {30},
number = {7},
pages = {917--929},
year = {2022},
issn = {1931-3128},
doi = {https://doi.org/10.1016/j.chom.2022.06.008},
url = {https://www.sciencedirect.com/science/article/pii/S1931312822003134},
author = {Eugene V. Koonin and Valerian V. Dolja and Mart Krupovic}
}

@article{kunin:2005,
    author = {Victor Kunin and Leon Goldovsky and Nikos Darzentas and Christos A. Ouzounis},
    title = {The net of life: Reconstructing the microbial phylogenetic network},
    journal = {Genome Res.},
    volume = {15},
    pages = {954--959},
    year = {2005}
}

@article{langley:1983,
    author = "C. H. Langley and J. F. Brookfield and N. Kaplan",
    title = "Transposable elements in mendelian populations. {I. A theory}",
    year = 1983,
    journal = "Genetics",
    volume = 104,
    pages = "457--571"
}

@article{lima-mendez:2017,
    author = "G. Lima-Mendez and J. {Van} Helden and A. Toussaint and R. Leplae",
    title = "Reticulate representation of evolutionary and functional relationships between phage genomes",
    year = 2008,
    journal = "Mol. Biol. Evol.",
    volume = 25,
    pages = "762--777"
}

@article{lobkovsky:2013,
    author = {Lobkovsky, Alexander E. and Wolf, Yuri I. and Koonin, Eugene V.},
    title = {Gene Frequency Distributions Reject a Neutral Model of Genome Evolution},
    journal = {Genome Biology and Evolution},
    volume = {5},
    number = {1},
    pages = {233-242},
    year = {2013},
    month = {01},
    issn = {1759-6653},
    doi = {10.1093/gbe/evt002},
    url = {https://doi.org/10.1093/gbe/evt002},
    eprint = {https://academic.oup.com/gbe/article-pdf/5/1/233/17919466/evt002.pdf},
}

@Article{manzano:2023,
author={Manzano-Morales, Saioa
and Liu, Yang
and Gonz{\'a}lez-Bod{\'i}, Sara
and Huerta-Cepas, Jaime
and Iranzo, Jaime},
title={Comparison of gene clustering criteria reveals intrinsic uncertainty in pangenome analyses},
journal={Genome Biol.},
year={2023},
month={Oct},
day={30},
volume={24},
number={1},
pages={250},
issn={1474-760X},
doi={10.1186/s13059-023-03089-3},
url={https://doi.org/10.1186/s13059-023-03089-3}
}

@article{manrubia:2002,
title = {At the Boundary between Biological and Cultural Evolution: The Origin of Surname Distributions},
journal = {J. Theor. Biol.},
volume = {216},
number = {4},
pages = {461-477},
year = {2002},
issn = {0022-5193},
doi = {https://doi.org/10.1006/jtbi.2002.3002},
url = {https://www.sciencedirect.com/science/article/pii/S002251930293002X},
author = {S. C. Manrubia and D. H. Zanette}
}

@article{moody:1988,
    author = "M. E. Moody",
    title = "A branching process model for the evolution of transposable elements",
    year = 1988,
    journal = "J. Math. Biol.",
    volume = 26,
    pages = "347--357"
}

@article{newman:2003,
author = {Newman, M. E. J.},
title = {The Structure and Function of Complex Networks},
journal = {SIAM Review},
volume = {45},
number = {2},
pages = {167-256},
year = {2003},
doi = {10.1137/S003614450342480},
URL = {https://doi.org/10.1137/S003614450342480},
eprint = {https://doi.org/10.1137/S003614450342480}
}

@book{newman:2010,
author = {Newman, M. E. J.},
year = {2010},
publisher = {Oxford University Press, New York},
title = {Networks: An introduction}
}

@article{pavlopoulos:2018,
    author = {Pavlopoulos, Georgios A and Kontou, Panagiota I and Pavlopoulou, Athanasia and Bouyioukos, Costas and Markou, Evripides and Bagos, Pantelis G},
    title = {Bipartite graphs in systems biology and medicine: a survey of methods and applications},
    journal = {GigaScience},
    volume = {7},
    number = {4},
    pages = {giy014},
    year = {2018},
    month = {02},
    issn = {2047-217X},
    doi = {10.1093/gigascience/giy014},
    url = {https://doi.org/10.1093/gigascience/giy014}
}

@Article{puigbo:2014,
author={Puigb{\`o}, Pere and Lobkovsky, Alexander E. and Kristensen, David M. and Wolf, Yuri I. and Koonin, Eugene V.},
title={Genomes in turmoil: quantification of genome dynamics in prokaryote supergenomes},
journal={BMC Biology},
year={2014},
month={Aug},
day={21},
volume={12},
number={1},
pages={66},
issn={1741-7007},
doi={10.1186/s12915-014-0066-4},
url={https://doi.org/10.1186/s12915-014-0066-4}
}

@article{reed:2004,
title = {A model explaining the size distribution of gene and protein families},
journal = {Mathematical Biosciences},
volume = {189},
number = {1},
pages = {97-102},
year = {2004},
issn = {0025-5564},
doi = {https://doi.org/10.1016/j.mbs.2003.11.002},
url = {https://www.sciencedirect.com/science/article/pii/S0025556403001962},
author = {William J Reed and Barry D Hughes}
}

@article{sela:2016,
    author = "I. Sela and Y. I. Wolf and E. V. Koonin",
    title = "Theory of prokaryotic genome evolution",
    year = 2016,
    journal = PNAS,
    volume = 113,
    pages = "11399--11407"
}

@article{snel:2002,
    author = "B. Snel and P. Bork and M. A. Huynen",
    title = "Genomes in flux: the evolution of archaeal and proteobacterial gene content",
    year = 2002,
    journal = {Genome Research},
    volume = 12,
    pages = "17--25"
}

@ARTICLE{simon:1955,
author = {Herbert Alexander Simon}, 
title = {On a class of skew distribution functions},
journal = {Biometrika},
volume = {42},
pages = {425--440},
year = {1955}
}

@Article{wolf:2016,
author={Wolf, Yuri I. and Makarova, Kira S. and Lobkovsky, Alexander E. and Koonin, Eugene V.},
title={Two fundamentally different classes of microbial genes},
journal={Nature Microbiology},
year={2016},
month={Nov},
day={07},
volume={2},
number={3},
pages={16208},
issn={2058-5276},
doi={10.1038/nmicrobiol.2016.208},
url={https://doi.org/10.1038/nmicrobiol.2016.208}
}

@article{wolf:2018,
    author = "Y. I. Wolf and D. Kazlauskas and J. Iranzo and A. Luc\'{\i}a-Sanz and J. H. Kuhn and M. Krupovic and V. V. Dolja and E. V. Koonin",
    title = "Origins and Evolution of the Global {RNA} Virome",
    year = 2018,
    journal = "mBio",
    volume = 9,
    pages = "e02329-18"
}

@article{yule:1925,
    author = "G. U. Yule",
    title = "{II.---A mathematical theory of evolution, based on the conclusions of Dr. J. C. Willis, F. R. S}",
    year = 1925,
    journal = PTRSB,
    volume = 213,
    pages = "21--87"
}

@article{yutin:2014,
title = {Origin of giant viruses from smaller DNA viruses not from a fourth domain of cellular life},
journal = {Virology},
volume = {466-467},
pages = {38-52},
year = {2014},
note = {Special issue: Giant Viruses},
issn = {0042-6822},
doi = {https://doi.org/10.1016/j.virol.2014.06.032},
url = {https://www.sciencedirect.com/science/article/pii/S0042682214003006},
author = {Natalya Yutin and Yuri I. Wolf and Eugene V. Koonin}
}

\end{document}